\documentclass[11pt]{article}
\newcommand{\Comment}[1]{{}}
\usepackage[textwidth = 430 pt, textheight = 630 pt]{geometry}
\usepackage{amssymb,euscript,amsmath,amsfonts}

\usepackage{color}
\definecolor{MyDarkBlue}{rgb}{0.15,0.15,0.45}
\usepackage[linktocpage=true]{hyperref}
\hypersetup{
colorlinks=true,
citecolor=MyDarkBlue,
linkcolor=MyDarkBlue,
urlcolor=MyDarkBlue,
pdfauthor={Neil Lambert },
pdftitle={ }, 
pdfsubject={hep-th}
}

\usepackage[numbers,sort&compress]{natbib}
\usepackage{hypernat}

\newcommand\ignore[1]{}
\def\one{{\,\hbox{1\kern-.8mm l}}}

\newcommand{\Cset}{{\,\,{{{^{_{\pmb{\mid}}}}\kern-.45em{\mathrm C}}}}}

\newcommand{\be}{\begin{equation}}
\newcommand{\ee}{\end{equation}}
\newcommand{\bea}{\begin{eqnarray}}
\newcommand{\eea}{\end{eqnarray}}

\numberwithin{equation}{section}
\begin{document}

\renewcommand{\thefootnote}{\fnsymbol{footnote}}

\rightline{MTH-KCL/2016-04}

   \vspace{1.8truecm}

 \centerline{\LARGE \bf {\sc  M2-Branes And The  (2,0) Superalgebra} }

\centerline{\LARGE \bf {\sc  }} \vspace{2truecm} \thispagestyle{empty} \centerline{
    {\large {\bf {\sc N.~Lambert$^{\,a,}$}\footnote{E-mail address: \href{mailto:neil.lambert@kcl.ac.uk}{\tt neil.lambert@kcl.ac.uk}} {\sc, \,\,D.~Sacco${}^{\,a,}$}}}\footnote{E-mail address: \href{mailto:damiano.sacco@kcl.ac.uk}{\tt damiano.sacco@kcl.ac.uk}}
  }

\vspace{1cm}
\centerline{${}^a${\it Department of Mathematics}}
\centerline{{\it King's College London, WC2R 2LS, UK}} 

\vspace{1.0truecm}

 
\thispagestyle{empty}

\centerline{\sc Abstract}
\vspace{0.4truecm}
\begin{center}
\begin{minipage}[c]{360pt}{
We present a generalization of the six-dimensional $(2,0)$ system of arXiv:1007.2982  to include a constant abelian 3-form. For vanishing  3-form this system is known to provide a variety descriptions of parallel M5-branes. For a particular choice of 3-form the system is shown to reduce to that of two M2-branes. Thus this generalised $(2,0)$ system provides a unified description of two  parallel M2-branes or M5-branes.
    \noindent }

\end{minipage}
\end{center}

\newpage 
 
\renewcommand{\thefootnote}{\arabic{footnote}}
\setcounter{footnote}{0}

\section{Introduction}

D$p$-branes are all related to each other in a straightforward way using T-duality which is valid microscopically in the open string description and also is manifest in the low energy Yang-Mills effective actions \cite{Taylor:1996ik}, although of course the quantum behaviour of these theories  drastically depends  on their dimension. Mathematically this occurs because all the maximally supersymmetric Yang-Mills theories are constructed by dimensional reduction of ten-dimensional super-Yang-Mills.

In M-theory there are M2-branes and M5-branes. While the field theories for multiple M2-branes are now known  \cite{Gustavsson:2007vu,Bagger:2007jr,Aharony:2008ug} (for a review see \cite{Bagger:2012jb}) the M5-brane remains  mysterious and believed to be non-lagrangian. Although there are various proposals for M5-brane dynamics that use a lagrangian but which require some specific limit to be taken  \cite{Aharony:1997th,Douglas:2010iu,Lambert:2010iw, Hull:2014cxa}. One still expects there to be some form of T-duality, inherited from string theory, that relates M5-branes to M2-branes. Even though there is no microscopic picture of these theories analogous to open strings one may still expect to see some universal structure in their field theory descriptions.  

One attempt to relate the M2-branes to M5-branes using T-duality was given in \cite{Jeon:2012fn}. The simple translational orbifold approach used in \cite{Taylor:1996ik} fails as translations are not a symmetry of the M2-brane Lagrangian. Nevertheless  the  modified approach of  \cite{Jeon:2012fn} leads from the periodic array of M2-branes to a variation of five-dimensional Super-Yang-Mills as a description of M5-branes.
 
In this paper we will generalise the six-dimensional (2,0) superalgebra construction of  \cite{Lambert:2010wm} by including a non-dynamical abelian background three-form.\footnote{Using such a three-form has also been considered by A. Gustavsson \cite{Gustavsson}.} Setting this to zero reproduces the previous results which have been proposed as a description of two M5-branes (here we specialise to the case of a positive definite Lie-3-algebra). In particular there is a covariantly constant vector which imposes constraints that require there to be an isometry along one direction which leads to five-dimensional super-Yang-Mills in the spacelike case \cite{Lambert:2010wm}, five-dimensional euclidean Super-Yang-Mills in the timelike case \cite{Hull:2014cxa} and quantum mechanics on instanton moduli space in the null case \cite{Lambert:2011gb}. These have all been argued to provide a description of the quantum $(2,0)$ theory \cite{Douglas:2010iu,Lambert:2010iw, Hull:2014cxa,Aharony:1997th}. We then show that turning on the background three-form allows some components of the vector to be dynamical but also forces a dimensional reduction to three dimensions leading to the maximally supersymmetric field theory of two M2-branes \cite{Gustavsson:2007vu,Bagger:2007jr}. Thus this generalized $(2,0)$ superalgebra provides a structure that contains aspects of both multiple M2-branes and M5-branes.

 There have also been papers which show that the maximally supersymmetric M2-brane Lagrangian   with a Nambu bracket for the 3-algebra leads to an abelian M5-brane \cite{Ho:2008nn,Ho:2008ve,Bandos:2008fr,Pasti:2009xc}. It might be possible to view the results here in a   complimentary context: starting from  the non-abelian $(2,0)$ superalgebra associated to multiple M5-branes and then obtaining M2-branes.

The structure of the paper is as follows. In section 2 we review the  non-abelian  (2,0) algebra of \cite{Lambert:2010wm} and the constraints on the fields that have to be satisfied for the algebra to close. We also briefly discuss how this algebra leads to various descriptions of M5-branes. In section 3 we propose a generalization of the  algebra  through the introduction of an abelian three-form $C_{\mu\nu\lambda}$, close the algebra and derive the constraints and equations of motion for the fields. In section 4 we find the central charges and the energy-momentum tensor associated to the generalized (2,0) algebra. In section 5 we relate our construction  to the maximally supersymmetric model describing two M2-branes and carry out the reduction. Section 5 has our conclusions.
 
\section{A (2,0) Algebra}

In \cite{Lambert:2010wm} a (2,0) algebra was realised on a non-abelian  six dimensional tensor multiplet. In particular the supersymmetry transformations from which the algebra arises were the following
\begin{align}
\delta X^i &= i\bar\epsilon \Gamma^i\Psi\nonumber\\
\delta Y^\mu & =0 \nonumber\\
\delta \Psi & = \Gamma^\mu\Gamma^i D_\mu X^i + \frac{1}{2\cdot 3!}H_{\mu\nu\lambda}\Gamma^{\mu\nu\lambda}\epsilon-\frac{1}{2}\Gamma_\mu\Gamma^{ij}[Y^\mu,X^i,X^j]\epsilon\nonumber\\
\delta H_{\mu\nu\lambda} & = 3i\bar\epsilon \Gamma_{[\mu\nu}D_{\lambda]}\Psi + i \bar\epsilon \Gamma^i\Gamma_{\mu\nu\lambda\rho}[Y^\rho,X^i,\Psi]\nonumber\\
\delta A_\mu(\cdot) & = i\bar\epsilon\Gamma_{\mu\nu}[Y^\nu,\Psi,\ \cdot\ ]\ ,
\end{align}
where $\Gamma^\mu,\Gamma^i$ are $32\times 32$ real $\Gamma$-matrices with $\mu,\nu,...=0,1,2,...,5$ and $i,j,...=6,7,8,9,10$. The spinors also satisfy
\begin{equation}
\Gamma_{012345}\epsilon = \epsilon\qquad \Gamma_{012345}\Psi = -\Psi\ ,
\end{equation}
and the three-form is self-dual:
\begin{equation}
H_{\mu\nu\lambda} = \frac{1}{3!}\varepsilon_{\mu\nu\lambda\rho\sigma\tau}H^{\rho\sigma\tau}\ .
\end{equation}
Note that the mass dimensions of the fields are
\begin{equation}
[X]=2\ ,\qquad [\Psi]=5/2\ ,\qquad [H] = 3\ ,\qquad [A] = 1\ ,\qquad [Y]=-1\ .
\end{equation}
The fields all take values in a Lie-3-algebra, that is in a vector space  endowed with a totally anti-symmetric product $[\,\,\,,\,\,\,,\,\,\,]$ from the vector space to itself. If we expand all in fields in terms of a basis for the 3-algebra  $\{T^a\}$, {\it i.e.} $X=X_a T^a$, then  
\begin{equation}
[X,Y,Z]_d=X_aY_bZ_c f^{abc}_{\quad \,\,d}\ ,
\end{equation}
where the structure constants of the 3-algebra $f^{abc}_{\quad \,\,d}$ are anti-symmetric in the upper indices.  Furthermore the triple product is required to satisfy the  fundamental identity which reads
\begin{equation}
[A,B,[X,Y,Z]] = [[A,B,X],Y,Z]+ [X,[A,B,Y],Z]+ [X,Y,[A,B,Z]]\ ,
\end{equation}
or equivalently, the structure constants need to satisfy:
\begin{equation}
f^{[abc}_{\quad \,\, e} f^{d]efg}=0\ .
\end{equation}
We also require the existence of a symmetric inner-product which is invariant under the action of the 3-algebra, which allows the definition of a metric structure
\begin{equation}
h^{ab}=\langle T^a,T^b\rangle\ .
\end{equation}
Which is equivalent to the condition $f^{[abcd]}=0$, where $f^{abcd}=f^{abc}{}_eh^{ed}$.

Derivatives on the fields are promoted to covariant derivatives with the introduction of a gauge field $A_\mu=(A_\mu)^b_{\,\,a}$ such that
\begin{equation}
D_\mu X_a=\partial_\mu X_a-(A_\mu)^b_{\,\,a} X_b=\partial_\mu X_a-A_\mu(X)_a\ ,
\end{equation}
and the field strength is defined as
 \begin{equation}
 F_{\mu\nu}\equiv-[D_\mu,D_\nu]\ .
 \end{equation}
 In \cite{Lambert:2010wm} it was shown that this algebra closes if the following set of equations of motion and constraints for the fields are satisfied
\begin{align}\label{oldeom}
0 =\quad& D^2X^i-\frac{i}{2}[Y^\mu,\bar \Psi,\Gamma_\mu\Gamma^i\Psi]-[Y^\mu,X^j,[Y_\mu,X^j,X^i]]\nonumber\\
0 =\quad&D_{[\mu}H_{\nu\lambda\rho}+\frac{1}{4}\varepsilon_{\mu\nu\lambda\rho\sigma\tau}[Y^\sigma,X^i,D^\tau X^i]+\frac{i}{8}\varepsilon_{\mu\nu\lambda\rho\sigma\tau}[Y^\sigma,\bar \Psi,\Gamma^\tau\Psi]\nonumber\\
0 =\quad& \Gamma^\mu D_\mu \Psi +\Gamma^\mu\Gamma^i[Y_\mu,X^i,\Psi]\nonumber\\
0 =\quad& F_{\mu\nu}(\cdot) - [Y^\lambda,H_{\mu\nu\lambda},\ \cdot\ ]\nonumber\\
0 =\quad&D_\mu Y^\nu=[Y^\mu,Y^\nu,\ \cdot\ ]=[Y^\mu, D_\mu (\cdot),\ \cdot'\ ]\ .
\end{align}
The aim of this paper is to generalise this algebra by including an abelian non-dynamical three-form $C_{\mu\nu\lambda}$ with mass dimension $[C]=-3$.   

Before doing so let us briefly recall how this system leads to various descriptions of M5-branes. For simplicity we consider the case of the four-dimensional 3-algebra $a=1,2,3,4$ with structure constants $f^{abcd}= \frac{2\pi }{k} \varepsilon^{abcd}$ and invariant metric $h^{ab} = \delta^{ab}$. The vector $Y^\mu$ is non-dynamical and can be fixed to the form
\begin{equation}
Y^\mu_a =  V^\mu\delta_a^4\ ,
\end{equation}
where we have fixed a particular direction in the 3-algebra and $V^\mu$ is a constant vector. Since all triple products involve $Y^\mu$ we see that the components of the fields along the $a=4$ direction decouple and become a free 6-dimensional abelian (2,0) multiplet which can be viewed as describing the centre of mass. 

Let us consider the interacting part consisting of fields with 3-algebra indices $\dot a=1,2,3$. The remaining constraints tell us that 
\begin{align}
V^\mu D_\mu (\ \cdot\ )_{\dot a} = 0 \ ,
\end{align}
so the interacting components of the fields cannot depend on the coordinate parallel to $V^\mu$. Furthermore we see that
\begin{equation}
F_{\mu\nu a}{}^{  b}=  V^\lambda f^{{ c}4 b}{}_{  a} H_{\mu\nu\lambda { c}}\ .
\end{equation}
In particular $V^\nu F_{\mu\nu a}{}^{  b}=0$ and we can  take $V^\mu A_{\mu a}{}^{  b}=0$ and $V^\mu \partial_\mu (\ \cdot\  )_{\dot a}=0$. 

To continue we must specify in what direction $V^\mu$ points. If it is spacelike then the $SO(1,5)$ Lorentz symmetry allows us to set $V^\mu = l \delta^\mu_5$ for some constant $l$. The resulting equations are then simply   those of maximally supersymmetric Yang-Mills with gauge group $SU(2)$   \cite{Lambert:2010wm}. Alternatively if $V^\mu$ is timelike then the $SO(1,5)$ Lorentz symmetry allows us to set $V^\mu = l \delta^\mu_0$. The resulting equations are now  those of maximally supersymmetric euclidean Yang-Mills theory with gauge group $SU(2)$  (and with an $SO(5)$ R-symmetry) \cite{Hull:2014cxa}. Finally if $V^\mu = l\delta^\mu_+$ is null then the self-duality constraint on $H_{\mu\nu\lambda}$ implies that $F_{\mu\nu}$ is self-dual (in the remaining four spatial directions). As such the ADHM construction can be used to solve for all the fields in terms of instanton moduli space \cite{Lambert:2011gb}. However the moduli are allowed to depend on $x^-$ and the resulting system can be quantized and leads to quantum mechanics (with $x^-$ as time) on instanton moduli space. All three of these descriptions have been proposed as giving the dynamics of multiple M5-branes  (see  \cite{Douglas:2010iu,Lambert:2010iw},\cite{Hull:2014cxa} and \cite{Aharony:1997th}   respectively).

 \section{Closure of the Algebra}
 
We consider the following extension of the (2,0) algebra
 \begin{align}\label{algebra}
 \delta X^i =& i\bar\epsilon \Gamma^i\Psi\nonumber\\
 \delta Y^\mu  =& \frac{i\alpha}{3!} \bar\epsilon \Gamma_{\lambda\rho}C^{\mu\lambda\rho}\Psi \nonumber\\  
 \delta \Psi  =& \Gamma^\mu\Gamma^i D_\mu X^i\epsilon + \frac{1}{2\cdot 3!}H_{\mu\nu\lambda}\Gamma^{\mu\nu\lambda}\epsilon\nonumber\\&-\frac{1}{2}\Gamma_\mu\Gamma^{ij}[Y^\mu,X^i,X^j]\epsilon +\frac{\beta}{3!} C_{\mu\nu\lambda}\Gamma^{\mu\nu\lambda}\Gamma^{ijk}[X^i,X^j,X^k] \epsilon\nonumber\\
 \delta H_{\mu\nu\lambda}  =& 3i\bar\epsilon \Gamma_{[\mu\nu}D_{\lambda]}\Psi + i \bar\epsilon \Gamma^i\Gamma_{\mu\nu\lambda\rho}[Y^\rho,X^i,\Psi]\nonumber\\&  +i\gamma \bar\epsilon (\star C)_{\mu\nu\lambda}\Gamma^{ij}[X^i,X^j,\Psi] + \frac{i\gamma'}{2}\bar\epsilon \Gamma_{[\mu\nu|\rho\sigma}C^{\rho\sigma}{}_{\lambda]}\Gamma^{ij}[X^i,X^j,\Psi] \nonumber\\
 \delta A_\mu(\cdot)  = &i\bar\epsilon\Gamma_{\mu\nu}[Y^\nu,\Psi,\ \cdot\ ] +\frac{i \delta }{3!} \bar\epsilon C^{\nu\lambda\rho}\Gamma_{\mu\nu\lambda\rho}\Gamma^i[X^i,\Psi,\ \cdot\ ]\ ,
 \end{align} 
 where $\alpha,\beta,\gamma,\gamma',\delta$ are constants to be determined and a dot ($\cdot$) denotes an arbitrary field.  There are additional terms that one could consider however the rationale behind this choice of algebra will become clear upon showing how a natural reduction to the M2-branes arises. In this section we will show that the superalgebra closes on shell and we will derive the equations of motion and the constraints that the fields need to satisfy. 
 
Before we consider the closure of the algebra we first observe that  
the fermion equation of motion can be obtained by imposing self-duality of $\delta H$. We find that
 \begin{align}
  \delta H_{\mu\nu\lambda}-(\star\delta  H)_{\mu\nu\lambda}  & = i\bar\epsilon\Gamma_{\mu\nu\lambda} (\Gamma^\rho D_\rho\Psi + \Gamma_\rho\Gamma^i[Y^\rho,X^i,\Psi] + \frac{\gamma}{3!}\Gamma_{\rho\sigma\tau}C^{\rho\sigma\tau}\Gamma^{ij}[X^i,X^j,\Psi])\ ,
 \end{align}
 provided that $\gamma'=3\gamma$ (otherwise one does not find a single expression on the right hand side). Thus we see that the Fermion equation of motion is
 \begin{align}\label{Psieom}
 \Gamma^\rho D_\rho\Psi + \Gamma_\rho\Gamma^i[Y^\rho,X^i,\Psi] + \frac{\gamma}{3!}\Gamma_{\rho\sigma\tau}C^{\rho\sigma\tau}\Gamma^{ij}[X^i,X^j,\Psi]=0\ .
 \end{align}
\subsection{Closure on $X^i$}
 We now proceed to close the algebra on the scalar fields $X^i$. We see that the algebra closes up to a translation and a gauge transformation, that is
 \begin{align}
 [\delta_1,\delta_2] X^i = v^\nu D_\nu X^i + \Lambda(X^i)\ ,
 \end{align}
 with  
 \begin{align}\label{gaugetr}
 v^\mu &=-2i(\bar\epsilon_2\Gamma^\mu \epsilon_1)\nonumber\\
 \Lambda(\cdot) &= -2i(\bar\epsilon_2\Gamma_{\lambda}\Gamma^i\epsilon_1)[Y^\lambda,X^i,\ \cdot\ ]-   i\beta(\bar\epsilon_2\Gamma_{\mu\nu\lambda}\Gamma^{jk}\epsilon_1)C^{\mu\nu\lambda}[X^j,X^k,\ \cdot\ ]\ .
 \end{align}
 We note that a new term, proportional to $C_{\mu\nu\lambda}$, now contributes to the definition of gauge transformation compared to the one defined in \cite{Lambert:2010wm}.
 \subsection{Closure on $Y^\mu$}
 Next we look at closing supersymmetry on $Y^\mu$. The expected form of the closure is 
 \begin{equation}
 [\delta_1,\delta_2] Y^\mu = v^\nu D_\nu Y^\mu + \Lambda(Y^\mu)\ ,
 \end{equation}
with $v^\mu$ and $\Lambda(\cdot)$ as defined in (\ref{gaugetr}).
Explicit calculation leads to 
 \begin{align}
 [\delta_1,\delta_2] Y^\mu  =& -\frac{i\alpha}{3} (\bar \epsilon_2\Gamma^\nu\epsilon_1)\,\, C^{\mu\lambda\rho}H_{\nu\lambda\rho}+\frac{2i\alpha}{3}\left(\bar\epsilon_2\Gamma_\nu\Gamma^i\epsilon_1\right) \,\,C^{\mu\nu\sigma}D_\sigma X^i
\nonumber \\
 &-\frac{i\alpha}{6}\left(\bar\epsilon_2 \Gamma_{\lambda\rho\sigma}\Gamma^{ij}\epsilon_1\right) C^{\mu\lambda\rho}[Y^\sigma,X^i,X^j]\nonumber\\
 &+\frac{i\alpha\beta}{3}\left(\bar \epsilon_2\Gamma_\lambda^{\,\,\,\tau\omega}\Gamma^{ijk}\epsilon_1\right)C^{\mu\lambda\rho} C_{\rho\tau\omega}[X^i,X^j,X^k]\ .
 \end{align}
 We see that imposing the constraint
 \begin{equation}\label{Cons1}
 D_\nu Y^\mu-\frac{\alpha}{6} \,C^{\mu\lambda\rho}H_{\nu\lambda\rho}=0\ ,
 \end{equation}
  turns the first term of the closure into a translation. Similarly, with the help of the constraint
 \begin{equation}\label{Cons2}
  C^{\mu\nu\sigma}D_\sigma X^i+\frac{3}{\alpha}\,[Y^\mu,Y^\nu,X^i]=0\ ,
 \end{equation}
 the second term of the closure represents the first part of a gauge transformation. We see that both these constraints are generalizations of ones found in \cite{Lambert:2010wm}, {\it c.f.}   (\ref{oldeom}).

 In order for the third line to turn into the part of a gauge transformation parametrized by $C^{\sigma\tau\omega}$ we need 
 \begin{equation}\label{Cons3}
 C^{\mu\lambda\rho} \left(\bar \epsilon_2\Gamma_{\lambda\rho\sigma}\Gamma^{ij}\epsilon_1\right)
 Y^\sigma= \frac{6\beta}{\alpha} C^{\sigma\tau\omega} \left(\bar \epsilon_2\Gamma_{\sigma\tau\omega}\Gamma^{ij}\epsilon_1\right)
 Y^\mu\ .
 \end{equation}
It is easily checked that if $\alpha=18\beta$ this condition is simply reduced to 
\begin{equation}
 C\wedge Y=0\ .
 \end{equation}
We will find that the condition $\alpha=18\beta$ also arises for closure on the other fields.

 We require the fourth term to vanish as it parametrizes neither a translation nor 
 a gauge transformation and hence\footnote{One might object that only the self-dual part of the left hand side must vanish but this possibility is eliminated by closure on $H_{\mu\nu\lambda}$.}
 \begin{equation}\label{Cons4}
 C_{ [\mu\nu}{}^\tau C_{\lambda]\tau}{}^\rho =0\ .
 \end{equation}
Note that this means that the components of $C_{\mu\nu\lambda}$ can be identified with the structure constants of a Lie-algebra. Since $\mu,\nu,...=0,1,2,...5$ this leads to only two possible choices: $su(2)$ and $so(4)=su(2)\oplus su(2)$.

 \subsection{Closure on $A_\mu$}
 From closing supersymmetry on the gauge field $A_\mu$ we expect to find
 \begin{align}
[\delta_1,\delta_2]A_\mu=-v^\nu F_{\mu\nu}+D_\mu\Lambda\ ,
 \end{align}
 Using the relations and constraints found so far, we find after some calculations that
 \begin{align}
  \left[\delta_1,\delta_2\right]A_\mu=&  2i (\bar \epsilon_2\Gamma^\nu\epsilon_1)\left([Y^\lambda,H_{\mu\nu\lambda},\ \cdot\ ]+\delta (\star C)_{\mu\nu\lambda}[X^i,D^\lambda X^i,\ \cdot\ ]  +\frac{i\delta}{2}(\star C)_{\mu\nu\lambda}[\bar{\Psi},\Gamma^\lambda\Psi,\ \cdot\ ]\right)\nonumber\\
  & +D_\mu \Lambda
  +2i \left(\bar\epsilon_2 \Gamma_\mu\Gamma^i\epsilon_1 \right)\left([Y^\nu,D_\nu X^i,\ \cdot\ ]-(\delta/6)C^{\sigma\tau\omega}[H_{\sigma\tau\omega},X^i,\ \cdot\ ]\right)\nonumber\\
   &+ 2 i(\beta +\delta/6)\left(\bar \epsilon_2\Gamma_{[\mu}^{\,\,\tau\omega}\Gamma^{ijk}\epsilon_1\right)C_{\nu]\tau\omega}[Y^\nu,[X^i,X^j,X^k],\ \cdot\ ]\nonumber\\
   &-i(\bar \epsilon_2\Gamma_{\mu\nu\sigma}\Gamma^{ij}\epsilon_1 )\left([Y^\nu,[Y^\sigma,X^i,X^j],\ \cdot\ ]+\frac{3\delta}{\alpha} [Y^\nu,[Y^\sigma,X^i,X^j],\ \cdot\ ]\right). 
 \end{align}
  We see that in order for the first term to represent a translation we must require the identification
   \begin{align}
 F_{\mu\nu}(\cdot) &=[Y^\lambda,H_{\mu\nu\lambda},\ \cdot\ ] + \delta(\star C)_{\mu\nu\lambda}[X^i,D^\lambda X^i, \ \cdot\ ] + \frac{i\delta}{2}(\star C)_{\mu\nu\lambda}[\bar\Psi,\Gamma^\lambda\Psi, \ \cdot\ ]\ ,
  \end{align}
  which generalizes the constraint in (\ref{oldeom}).  By looking at the form the closure needs to take, we require the last three terms to vanish. This imposes the correction to the known constraint
  \begin{equation}
  [Y^\nu,D_\nu X^i,\ \cdot\ ]-\frac{\delta}{6}\,C^{\sigma\tau\omega}[H_{\sigma\tau\omega},X^i,\ \cdot\ ]=0\ ,
  \end{equation}
as well as the relations between the coefficients
\begin{equation}
\delta=-6\beta,\quad\qquad \alpha=-3\delta\ .
\end{equation}

 \subsection{Closure on $H_{\mu\nu\lambda}$}
 
 Closing the algebra on $H_{\mu\nu\lambda}$ is somewhat more lengthy, and in the process we found the Mathematica GAMMA package quite helpful \cite{Gran:2001yh}. Supersymmetry should close up to a translation and a gauge transformation
 \begin{equation}
 [\delta_1,\delta_2]H_{\mu\nu\lambda}  = v^\rho D_\rho H_{\mu\nu\lambda}  + \Lambda(H_{\mu\nu\lambda})\ .
 \end{equation}
Since the calculation is quite involved we will not provide the full details here. Rather we note that  in order to close the algebra numerous terms are required to vanish as they parametrize neither a translation, nor a gauge transformation. This is the case if the following relations among the coefficients hold
\begin{equation}\label{relcoeff}
\gamma'=\,\,3\gamma,\qquad 
\gamma'=\,\,9\beta,\qquad 
\delta=-2\gamma\ .
\end{equation}
Then the remaining terms, making use of the constraints found so far, take the form
 \begin{align}
 \left[\delta_1,\delta_2\right]H_{\mu\nu\lambda}=&v^\rho D_\rho H_{\mu\nu\lambda}-2i(\bar \epsilon_2\Gamma_\sigma\Gamma^i\epsilon_1)[Y^\sigma,X^i, H_{\mu\nu\lambda}]\nonumber\\
 &-i\beta\left(\bar \epsilon_2 \Gamma_{\sigma\tau\omega}\Gamma^{ij}\epsilon_1   \right)C^{\sigma\tau\omega}[X^i,X^j,H_{\mu\nu\lambda }]\nonumber \\
 &+4 v^\rho \biggl( D_{[\lambda}H_{\mu\nu\rho]} +  \frac{1}{4}\varepsilon_{\mu\nu\lambda\rho\sigma\tau}[Y^\sigma,X^i,D^\tau X^i]- \gamma(\star C)_{[\mu\nu\lambda}[X^i,X^j,[Y_{\rho]},X^i,X^j]]\nonumber\\
 &  + \frac{i}{8}\varepsilon_{\mu\nu\lambda\rho\sigma\tau}[Y^\sigma,\bar\Psi,\Gamma^{\tau}\Psi]- i\gamma(\star C)_{[\mu\nu\lambda}[X^i,\bar\Psi,\Gamma_{\rho]}\Gamma^i\Psi]\biggr)\ ,
 \end{align}
 We see that the first three terms represent a translation and a gauge transformation.
The algebra then closes on shell and we find  the equation of motion for $H_{\mu\nu\lambda}$
 \begin{align}\label{Heom}
 D_{[\lambda}H_{\mu\nu\rho]} &= - \frac{1}{4}\varepsilon_{\mu\nu\lambda\rho\sigma\tau}[Y^\sigma,X^i,D^\tau X^i]+ \gamma(\star C)_{[\mu\nu\lambda}[X^i,X^j,[Y_{\rho]},X^i,X^j]]\nonumber\\
 &  - \frac{i}{8}\varepsilon_{\mu\nu\lambda\rho\sigma\tau}[Y^\sigma,\bar\Psi,\Gamma^{\tau}\Psi] + i\gamma (\star C)_{[\mu\nu\lambda}[X^i,\bar\Psi,\Gamma_{\rho]}\Gamma^i\Psi]\ .
 \end{align} 
 
 \subsection{Closure on $\Psi$}
 
 Closure of supersymmetry on the fermion $\Psi$ should be obtained up to a translation and a gauge transformation
 \begin{equation}
 [\delta_1,\delta_2]\Psi = v^\rho D_\rho \Psi  + \Lambda(\Psi)\ .
 \end{equation}
An explicit calculation, making use of the Gamma package \cite{Gran:2001yh} and the constraints found so far, gives
\begin{align}
[\delta_1,\delta_2]\Psi =& v^\rho D_\rho \Psi  + \Lambda(\Psi)\nonumber\\
&+\frac{3i}{4}(\bar \epsilon_2\Gamma_\sigma\epsilon_1)\Gamma^\sigma\left( \Gamma^\rho D_\rho\Psi + \Gamma_\rho\Gamma^i[Y^\rho,X^i,\Psi] + \frac{\gamma}{3!}\Gamma_{\rho\sigma\tau}C^{\rho\sigma\tau}\Gamma^{ij}[X^i,X^j,\Psi]\right)\nonumber\\
&-\frac{i}{4}(\bar \epsilon_2\Gamma_\sigma\Gamma^j\epsilon_1)\Gamma^\sigma\Gamma^j\left( \Gamma^\rho D_\rho\Psi + \Gamma_\rho\Gamma^i[Y^\rho,X^i,\Psi] + \frac{\gamma}{3!}\Gamma_{\rho\sigma\tau}C^{\rho\sigma\tau}\Gamma^{ij}[X^i,X^j,\Psi]\right)\ .
\end{align}
We see that in order to close the algebra the terms other than the translation and the gauge transformation need to vanish. This is achieved upon imposing the Fermion equation of motion, which agrees with (\ref{Psieom}).

\subsection{Bosonic Equations of Motion}

We can vary the Fermion equation of motion (\ref{Psieom}) to find the equations of motion for $X^i$ and $H_{\mu\nu\lambda}$. We find, making use of the constraints found so far, the following variation
\begin{align}
& \Big(D^2 X^i-\frac{i}{2}[Y^\sigma,\bar \Psi,\Gamma_\sigma\Gamma^i\Psi]+[Y^\sigma,X^j,[Y_\sigma,X^j,X^i]]\nonumber\\&+\frac{i\gamma}{3!}C^{\sigma\tau\omega}[\bar \Psi,\Gamma_{\sigma\tau\omega}\Gamma^{ij}\Psi,X^j]+\beta\gamma C^{\sigma\tau\omega}C_{\sigma\tau\omega}[[X^i,X^j,X^k],X^j,X^k]\Big)\Gamma^i\epsilon\nonumber\\
& +\frac{1}{3!} \Big( D_{\mu}H_{\nu\lambda\rho}+  \frac{1}{4}\varepsilon_{\mu\nu\lambda\rho\sigma\tau}[Y^\sigma,X^i,D^\tau X^i]- \gamma(\star C)_{\mu\nu\lambda}[X^i,X^j,[Y_{\rho},X^i,X^j]]\nonumber\\
&  + \frac{i}{8}\varepsilon_{\mu\nu\lambda\rho\sigma\tau}[Y^\sigma,\bar\Psi,\Gamma^{\tau}\Psi] - i\gamma (\star C)_{\mu\nu\lambda}[X^i,\bar\Psi,\Gamma_{\rho}\Gamma^i\Psi] \Big)\Gamma^{\mu\nu\lambda\rho}\epsilon=0\ .
\end{align}
We see that the equation of motion for $H_{\mu\nu\lambda}$ agrees with the one found by requiring closure of the algebra (\ref{Heom}). Moreover, we find the equation of motion for $X^i$
  \begin{align}\label{HXeom}
  	D^2 X^i&=\frac{i}{2}[Y^\sigma,\bar \Psi,\Gamma_\sigma\Gamma^i\Psi]-[Y^\sigma,X^j,[Y_\sigma,X^j,X^i]]\nonumber\\&-\frac{i\gamma}{3!}C^{\sigma\tau\omega}[\bar \Psi,\Gamma_{\sigma\tau\omega}\Gamma^{ij}\Psi,X^j]-\beta\gamma C^{\sigma\tau\omega}C_{\sigma\tau\omega}[[X^i,X^j,X^k],X^j,X^k]\ .
  \end{align}
Therefore we have determined the equations of motion for all the degrees of freedom of the (2,0) tensor multiplet.

 \subsection{Summary}

 We have shown that the (2,0) algebra (\ref{algebra}) we proposed closes on shell. We found corrections to the equations of motion and constraints (\ref{oldeom}), which we list here for convenience.  Since we are free to rescale $C_{\mu\nu\lambda}$ we can, without loss of generality, set the coefficients of the (2,0) algebra to the specific values
 \begin{equation}
 \alpha=3\,\,\quad\beta=1/3!\,\,\quad \gamma=1/2\,\,\quad\delta=-1\,\,\quad\gamma'=3/2\ ,
 \end{equation}
 which respect the relations found in the closure of the algebra.
 The equations of motion for the fields of the tensor multiplet are
 \begin{align}\label{eomfixed}
  0&=D^2 X^i-\frac{i}{2}[Y^\sigma,\bar \Psi,\Gamma_\sigma\Gamma^i\Psi]+[Y^\sigma,X^j,[Y_\sigma,X^j,X^i]]\nonumber\\&+\frac{i}{2\cdot3!}C^{\sigma\tau\omega}[\bar \Psi,\Gamma_{\sigma\tau\omega}\Gamma^{ij}\Psi,X^j]+\frac{1}{2\cdot 3! } C^{\sigma\tau\omega}C_{\sigma\tau\omega}[[X^i,X^j,X^k],X^j,X^k]\nonumber\\
 0 &= D_{[\lambda}H_{\mu\nu\rho]}+  \frac{1}{4}\varepsilon_{\mu\nu\lambda\rho\sigma\tau}[Y^\sigma,X^i,D^\tau X^i]- \frac{1}{2}(\star C)_{[\mu\nu\lambda}[X^i,X^j,[Y_{\rho]},X^i,X^j]]\nonumber\\
  &  + \frac{i}{8}\varepsilon_{\mu\nu\lambda\rho\sigma\tau}[Y^\sigma,\bar\Psi,\Gamma^{\tau}\Psi] - \frac{i}{2} (\star C)_{[\mu\nu\lambda}[X^i,\bar\Psi,\Gamma_{\rho]}\Gamma^i\Psi]\nonumber \\
 0&= \Gamma^\rho D_\rho\Psi + \Gamma_\rho\Gamma^i[Y^\rho,X^i,\Psi] + \frac{1}{2\cdot3!}\Gamma_{\rho\sigma\tau}C^{\rho\sigma\tau}\Gamma^{ij}[X^i,X^j,\Psi]\ ,
 \end{align}
 while the additional constraints for the algebra to close on shell are
 \begin{align}\label{Constraintsfixed}
 0&= F_{\mu\nu}(\cdot) -[Y^\lambda,H_{\mu\nu\lambda},\ \cdot\ ] +(\star C)_{\mu\nu\lambda}[X^i,D^\lambda X^i, \ \cdot\ ] + \frac{i}{2}(\star C)_{\mu\nu\lambda}[\bar\Psi,\Gamma^\lambda\Psi, \ \cdot\ ]\nonumber\\
    0&=D_\nu Y^\mu-\frac{1}{2} \,C^{\mu\lambda\rho}H_{\nu\lambda\rho}\nonumber\\
    0&=C^{\mu\nu\sigma}D_\sigma (\cdot)+\,[Y^\mu,Y^\nu,\ \cdot\ ]\nonumber\\
      0&=[Y^\nu,D_\nu \cdot\ ,\ \cdot'\ ]+\frac{1}{3!}\,C^{\sigma\tau\omega}[H_{\sigma\tau\omega},\ \cdot\ ,\ \cdot'\ ]\nonumber\\
      0&=C\wedge Y\ .
 \end{align}
 Note that using the second constraint the fourth constraint can be rewritten as
 \begin{align}
 0 =[Y^\nu,D_\nu \cdot\ ,\ \cdot' ]+\frac{1}{3} [D_\nu Y^\nu,\ \cdot\ ,\ \cdot'\ ]\ .
 \end{align}

 The equations of motion (\ref{eomfixed}) are invariant under the (2,0) supersymmetry realised by the variations
  \begin{align}\label{algebrafixed}
  \delta X^i &= i\bar\epsilon \Gamma^i\Psi\nonumber\\
  \delta Y^\mu  &= \frac{i}{2} \bar\epsilon \Gamma_{\lambda\rho}C^{\mu\lambda\rho}\Psi \nonumber\\  
  \delta \Psi  &= \Gamma^\mu\Gamma^i D_\mu X^i\epsilon + \frac{1}{2\cdot 3!}H_{\mu\nu\lambda}\Gamma^{\mu\nu\lambda}\epsilon\nonumber\\&-\frac{1}{2}\Gamma_\mu\Gamma^{ij}[Y^\mu,X^i,X^j]\epsilon +\frac{1}{3!^2} C_{\mu\nu\lambda}\Gamma^{\mu\nu\lambda}\Gamma^{ijk}[X^i,X^j,X^k] \epsilon\nonumber\\
  \delta H_{\mu\nu\lambda}  &= 3i\bar\epsilon \Gamma_{[\mu\nu}D_{\lambda]}\Psi + i \bar\epsilon \Gamma^i\Gamma_{\mu\nu\lambda\rho}[Y^\rho,X^i,\Psi]\nonumber\\&  +\frac{i}{2} \bar\epsilon (\star C)_{\mu\nu\lambda}\Gamma^{ij}[X^i,X^j,\Psi] + \frac{3i}{4}\bar\epsilon \Gamma_{[\mu\nu|\rho\sigma}C^{\rho\sigma}{}_{\lambda]}\Gamma^{ij}[X^i,X^j,\Psi] \nonumber\\
 \delta A_\mu(\cdot)  &= i\bar\epsilon\Gamma_{\mu\nu}[Y^\nu,\Psi,\ \cdot\ ] -\frac{i  }{3!} \bar\epsilon C^{\nu\lambda\rho}\Gamma_{\mu\nu\lambda\rho}\Gamma^i[X^i,\Psi,\ \cdot\ ]\ .
 \end{align}  
 
 \section{Conserved Currents}

In this section we construct  the supercurrent $S^\mu$ and  energy-momentum tensor $T_{\mu\nu}$ associated to the supersymmetry algebra realised in (\ref{algebra}). We can then deduce
the form of the superalgebra including the central charges.  

 The supercurrent can be easily computed by  
  \begin{align}
  \bar \epsilon S^\mu = 2\pi i\langle \overline{\delta_\epsilon\Psi},\Gamma^\mu\Psi\rangle \ .
  \end{align}
Note  the pre-factor of $2\pi$ which is needed to produce the correct energy-momentum tensor and will be justified in due course. Explicitly we find
 \begin{align}
S^\mu=&-2\pi i\langle D_\nu X^i ,\Gamma^\nu\Gamma^i\Gamma^\mu \Psi\rangle+\frac{\pi i}{3!} \langle H_{\sigma\tau\omega},\Gamma^{\sigma\tau\omega}\Gamma^\mu\Psi\rangle-\pi i\langle[Y_\nu,X^i,X^j],\Gamma^\nu\Gamma^{ij}\Gamma^\mu\Psi\rangle\nonumber\\ &+\frac{\pi i}{3\cdot 3!}C_{\sigma\tau\omega}\langle[X^i,X^j,X^k],\Gamma^{ijk}\Gamma^{\sigma\tau\omega}\Gamma^\mu\Psi\rangle\ .
 \end{align}
The supercurrent is indeed found to be conserved on shell. 

 Next we construct the energy-momentum tensor, which after some trial and error, reads
\begin{align}
T_{\mu\nu}=&2\pi\langle D_\mu X^i,D_\nu X^i\rangle-\pi\eta_{\mu\nu} \langle D_\lambda X^i,D^\lambda X^i\rangle+\pi\langle[X^i,X^j,Y_\mu],[X^i,X^j,Y_\nu]\rangle\nonumber\\
-&\frac{\pi}{2}\eta_{\mu\nu}\langle[X^i,X^j,Y_\lambda],[X^i,X^j,Y^\lambda]\rangle+\frac{\pi}{2}\langle H_{\mu\lambda\rho},H_\nu^{\,\,\lambda\rho}\rangle-i\pi\langle \bar \Psi,\Gamma_\mu D_\nu\Psi\rangle\nonumber\\
-&i\pi\langle \bar \Psi,\Gamma_\nu D_\mu\Psi\rangle+i\pi\eta_{\mu\nu}\langle\bar \Psi,\Gamma^\lambda D_\lambda \Psi\rangle-i\pi\eta_{\mu\nu}\langle[\bar \Psi,Y^\lambda,X^i],\Gamma_\lambda\Gamma^i\Psi\rangle\nonumber\\
&+\frac{\pi}{3!}\langle[X^i,X^j,X^k],[X^i,X^j,X^k]\rangle( \,C_{\mu\tau\omega}C_\nu^{\,\,\,\,\tau\omega}-\frac{1}{3!}\eta_{\mu\nu}C^2)\nonumber\\
&+\frac{\pi}{3!}C_{\mu\lambda\rho}(\star C)_{\nu}{}^{\lambda\rho}\langle[X^i,X^j,X^k],[X^i,X^j,X^k]\rangle-\frac{i\pi }{ 3!}\eta_{\mu\nu}C^{\sigma\tau\omega}\langle[\bar{\Psi},\Gamma_{\sigma\tau\omega}\Gamma^{ij}\psi,X^i ],X^{j}\rangle\ .
\end{align}
 The energy-momentum tensor is found to satisfy $\partial^\mu T_{\mu \nu}=0$ using the equations of motion and constraints for the fields derived in the previous section.\footnote{In fact conservation allows for arbitrary coefficients of the $C_{\mu\lambda\rho} C_{\nu}{}^{\lambda\rho}$ and $C_{\mu\lambda\rho}(\star C)_{\nu}{}^{\lambda\rho}$ terms that we have fixed by considering the super-algebra below.}  Although we note that the bosonic part is not symmetric for a general choice of three-form due to the $C_{\mu\lambda\rho}(\star C)_{\nu}{}^{\lambda\rho}$ term (as well as the more familiar asymmetry arising from the fermions). The $2\pi$ pre-factor was justified  in \cite{Maldacena:1997de} to agree with charge quantization and also in  \cite{Hull:2014cxa} to reproduce the correct energy density for M2-branes ending on M5-branes. It  also  leads to the correct matching of instanton-solitons with KK tower modes \cite{Hull:2014cxa}.

 In order to derive the super-algebra we make use of the the chain of identities
 \begin{equation}
 i\bar \epsilon^{B}\{Q_A,Q_B \}=i\{\bar \epsilon Q,Q_A \}=\delta_\epsilon Q_A=\int d^5 x \,\,(\delta_\epsilon S^0)_A\ ,
 \end{equation}
 where
 \begin{equation}
 Q=\int d^5x\,\, S^0\ .
 \end{equation}
 Since by construction $\{Q_A,Q_B\}$ is symmetric in $A,B$, we can extract the momentum 
 \begin{equation}
P_\nu = \int d^5x T_{0\nu}\ ,
 \end{equation} and the central charges ($Z^i_\mu, Z^{ij}_{\mu\nu\lambda})$ following the expansion
 \begin{equation}
 \{Q_A,Q_B\}=2 (\Gamma^\mu C^{-1})_{AB}P_\mu+(
 \Gamma^\mu\Gamma^iC^{-1})_{AB}Z^i_\mu+\frac{1}{2!\cdot 3!}(\Gamma^{\mu\nu\lambda}\Gamma^{ij}C^{-1})_{AB}Z^{ij}_{\mu\nu\lambda}\ .
 \end{equation}
In case of vanishing Fermions, we find the following central charges. For $Z^i_\mu$ we find  
 \begin{align}
Z_0^i=&4\pi\int d^5x\  \langle[Y_0,X^i,X^j],D^0X^j\rangle-\langle[Y_{\dot \mu},X^i,X^j],D^{\dot \mu}X^j\rangle \\
Z^i_{\dot  \mu}=&4\pi\int d^5x\  \langle[Y^0,X^i,X^j],D_{\dot\mu} X^j\rangle+\langle[Y_{\dot \mu},X^i,X^j],D^0 X^j\rangle \nonumber\\
&\qquad \qquad+\langle H_{0\dot \mu\dot \nu},D^{\dot \nu}X^i\rangle+\frac{1}{3}C^+_{0\dot\mu\dot\nu}\langle[X^j,X^k,X^l],D^{\dot \nu} X^{m}\rangle\varepsilon^{ijklm} \nonumber\\
&\qquad\qquad -  C^+_{0\dot\mu\dot\nu}\langle[X^i,X^j,X^k],[Y^{\dot\nu},X^j,X^k]\rangle\ ,
\end{align}
 while $Z^{ij}_{\mu\nu\lambda}$ reads (all the expressions should be taken to be anti-symmeterized in $i,j$ and $\dot\mu,\dot\nu,\dot\lambda$ where  dotted indices only run  over spatial coordinates $\dot \mu,\dot \nu=1,2,\dots,5$.)
\begin{align}
Z^{ij}_{0\dot\mu\dot\nu}=& { 4 \pi}  \int d^5 x\  2\langle[Y_{\dot \mu},X^i,X^k],[Y_{\dot \nu},X^k,X^j]\rangle-\langle[Y_{\dot \nu},X^k,X^l],D_{\dot \mu}X^m\rangle\varepsilon^{ijklm} \nonumber\\
&\qquad\qquad+  \frac 12\langle H_{0\dot \mu\dot \nu},[Y^0,X^i,X^j]\rangle-\frac 12\langle H_{\dot \mu\dot \nu\dot \rho},[Y^{\dot \rho},X^i,X^j]\rangle- 2\langle D_{\dot \mu}X^i,D_{\dot \nu}X^j\rangle\nonumber\\
&\qquad\qquad-  \langle( C_{\dot \mu\dot \nu\dot \rho} D^{\dot \rho} X^k+C_{0\dot\mu\dot\nu}D_0 X^k),[X^i,X^j,X^k]\rangle \nonumber\\
&\qquad\qquad+  \frac12\langle (C_{\dot \mu\dot \nu\dot \rho} [Y^{\dot\rho},X^k,X^n]-C_{0\dot\mu\dot\nu}[Y^0,X^k,X^n]),[X^l,X^m,X^n]\rangle\varepsilon^{ijklm} 
\nonumber\\
&\qquad\qquad- \frac{1}{2\cdot 3!}  \langle [X^k,X^l,X^m],(2 C_{0\dot\nu\dot\rho}H_{0\dot\mu}^{\,\,\,\,\,\,\dot \rho } + C_{\dot\nu\dot\rho\dot\sigma}H_{\dot\mu}^{\,\,\dot\rho\dot\sigma})\rangle\varepsilon^{ijklm} \\
Z^{ij}_{\dot \mu\dot \nu\dot \lambda}=& {4 \pi} \int d^5 x\ \frac{1}{2} \langle H_{\dot \mu\dot \nu\dot \lambda},[Y^0,X^i,X^j]\rangle-\frac{3}{2}\langle H_{0\dot \mu \dot \nu},[Y_{\dot \lambda},X^i,X^j]\rangle   \nonumber\\
&\qquad\qquad-  \langle( C_{\dot \mu\dot \nu\dot \lambda}D_0X^{k}+3 C_{0\dot \mu\dot \nu}D_{\dot \lambda} X^k),[X^i,X^j,X^k]\rangle \nonumber\\
&\qquad\qquad-\frac12 \langle( C_{\dot \mu\dot \nu\dot \lambda} [Y_0,X^m,X^n]+3 C_{0\dot \mu\dot \nu}[Y_{\dot \lambda},X^m,X^n]),[X^k,X^l,X^n]\rangle\varepsilon^{ijklm} \nonumber\\
&\qquad\qquad+\,\,\frac{1}{4} \langle(  C_{\dot \mu\dot \nu\dot \rho}H_{0\dot \lambda}^{\,\,\,\,\,\dot \rho}-C_{0\dot \lambda\dot \rho}H_{\dot \mu\dot \nu}^{\,\,\,\,\,\dot \rho}),[X^k,X^l,X^m]\rangle\varepsilon^{ijklm}\ .
\end{align}

\section{From (2,0) to 2 M2's}

As recalled in section 2 previous work has examined the dynamical systems that arise from the above system when $C_{\mu\nu\lambda}$ vanishes  \cite{Lambert:2010wm,Hull:2014cxa,Lambert:2011gb}. To this end let us split up spacetime into the directions $\alpha,\beta=0,1,2$ and $a,b=3,4,5$ and fix  
\begin{equation}
 C_{abc}=l^3\,\varepsilon_{abc}\ ,
 \end{equation}
  where $l$ has dimension of length. 
This breaks to the $SO(1,5)$ Lorentz symmetry to $SO(1,2)\times SO(3)$. We will see that this $SO(3)$  enhances the $SO(5)$ R-symmetry to $SO(8)$.

Recall the constraints found upon closing the (2,0) algebra (\ref{algebra}) on the tensor multiplet
\begin{align}\label{Constraints}
 0&= F_{\mu\nu}(\cdot) -[Y^\lambda,H_{\mu\nu\lambda},\ \cdot\ ] +(\star C)_{\mu\nu\lambda}[X^i,D^\lambda X^i, \ \cdot\ ] +  \frac{i}{2}(\star C)_{\mu\nu\lambda}[\bar\Psi,\Gamma^\lambda\Psi, \ \cdot\ ]\nonumber\\
 0&=D_\nu Y^\mu-\frac{1}{2} \,C^{\mu\lambda\rho}H_{\nu\lambda\rho}\nonumber\\
 0&=C^{\mu\nu\sigma}D_\sigma (\cdot)+\,[Y^\mu,Y^\nu,\ \cdot\ ]\nonumber\\
 0&=[Y^\nu,D_\nu \cdot,\cdot']+\frac{1}{3!}\,C^{\sigma\tau\omega}[H_{\sigma\tau\omega},\ \cdot\ ,\ \cdot'\ ]\ .
\end{align}
We now look at the third constraint
 \begin{equation}
 C^{\mu\nu\sigma}D_\sigma ( \cdot )+\,[Y^\mu,Y^\nu,\ \cdot\ ]=0\ ,
 \end{equation}
The simplest way to solve this constraint is to take the fields independent of the the $x^a$ spatial directions: $\partial_a(\cdot)=0$. Then the constraint is solved for 
 \begin{equation}
 A_a(\cdot)  =\frac{1}{2l^{3}}\varepsilon_{abc}[Y^b,Y^c,\ \cdot\ ]\ .
 \end{equation}
 Next we look at the last constraint 
 \begin{equation}
  [Y^\nu,D_\nu \ \cdot\ ,\ \cdot'\ ]+\frac{1}{6}\,C^{\sigma\tau\omega}[H_{\sigma\tau\omega},\ \cdot\ ,\ \cdot'\ ]=0\ ,
 \end{equation}
 and we see that a solution is given by 
 \begin{equation}\label{constsol}
 Y^\alpha=0\qquad  H_{abc}=-\frac{1}{ l^6} [Y_a,Y_b,Y_c]\ ,
 \end{equation}
 where to obtain the last relation we used the fundamental identity. Note that the second constraint is also solved by (\ref{constsol}). Finally  the first constraint is satisfied if in addition we have 
 \begin{equation}
 H_{\alpha ab}=\frac{1}{ l^3}\varepsilon_{abc}D_{\alpha}Y^c\ .
 \end{equation}
We note that similar expressions for $H_{\mu\nu\lambda}$ appeared in \cite{Ho:2008ve}. 
We also find that 
\begin{equation}
  F_{\alpha\beta}(\cdot) =-\frac{1}{l^3}\varepsilon_{\alpha\beta\gamma}[Y_a,D^\gamma Y^a,\ \cdot\ ] -l^3\varepsilon_{\alpha\beta\gamma}[X^i,D^\gamma X^i, \ \cdot\ ] - \frac{il^3}{2}\varepsilon_{\alpha\beta\gamma}[\bar\Psi,\Gamma^\gamma\Psi, \ \cdot\ ]\ .
\end{equation}
 To summarise, we found a solution to the constraints (\ref{Constraints}) given by
\begin{align}\label{solconst} 
\partial_a(\cdot)&=Y^\alpha=0\nonumber\\
 A_a(\cdot)  &=\frac{1}{2 l^{3}}\varepsilon_{abc}[Y^b,Y^c,\ \cdot\ ]\nonumber\\   
    F_{\alpha\beta}(\cdot) &=-\frac{1}{l^3}\varepsilon_{\alpha\beta\gamma}[Y_a,D^\gamma Y^a,\ \cdot\ ] -l^3\varepsilon_{\alpha\beta\gamma}[X^i,D^\gamma X^i, \ \cdot\ ] - \frac{il^3}{2}\varepsilon_{\alpha\beta\gamma}[\bar\Psi,\Gamma^\gamma\Psi, \ \cdot\ ]\nonumber\\ 
    H_{abc}&=-\frac{1}{ l^6} [Y_a,Y_b,Y_c] \nonumber\\ H_{\alpha ab}&=\frac{1}{ l^3}\varepsilon_{abc}D_{\alpha}Y^c\ ,
\end{align}
with the other components of $H_{\mu\nu\lambda}$ fixed by self-duality.
We now wish to implement   the solution to the constraints that we found into the algebra (\ref{algebrafixed}). We see that since the fields are required to be independent of the three spatial directions, a dimensional reduction naturally arises.

Let us now look at the supersymmetry transformations and apply the solution to the constraints (\ref{solconst}). We find, noting that the fields now depend only on $x^\alpha$, for the fermions
\begin{align}
\delta\Psi &= \Gamma^\alpha\Gamma^iD_\alpha X^i\epsilon+\frac{1}{2 l^3}\Gamma^{ab}\Gamma_{345}\Gamma^i[Y^a,Y^b,X^i]\epsilon - \frac{1}{3!l^6}\Gamma_{abc}[Y^a,Y^b,Y^c] \epsilon\nonumber\\&+ \frac{1}{ l^3}\Gamma^\alpha \Gamma^{c}\Gamma_{345}D_\alpha Y^c\epsilon - \frac{1}{2}\Gamma^{a}\Gamma^{ij}[Y^a,X^i,X^j] \epsilon + \frac{1}{3!l^3}\Gamma_{345}\Gamma^{ijk}[X^i,X^j,X^k]\epsilon\ ,
\end{align}
and for the bosons
\begin{align}
\delta X^i &= i\bar\epsilon\Gamma^i\Psi\nonumber\\
\delta Y^a & =  i l^3\bar\epsilon  \Gamma^{a}\Gamma_{345}\Psi\nonumber\\
\delta A_\alpha (\cdot )& = i\bar\epsilon\Gamma_{\alpha}\Gamma^b[Y^b,\Psi,\ \cdot\ ] - i l^3\bar\epsilon\Gamma_\alpha\Gamma_{345}\Gamma^i[X^i,\Psi,\ \cdot\ ]\ .
\end{align}
We can now discuss how the degrees of freedom of the two theories are related. The eight scalars parametrizing  fluctuations in the directions transverse to the M2-branes worldvolume will consist of the five scalars $X^i$ of the (2,0) tensor multiplet and the three remaining scalars $Y^\alpha$. Therefore we can  define the three-dimensional scalars:
\begin{equation}
X^I\equiv(l^{-3/2}Y^a,l^{3/2}X^i)\ ,
\end{equation}
where now $I,J=3,4,5,...,10$. Note that no other bosonic degrees of freedom are present since $H_{\mu\nu\lambda}$ is fixed by the constraints (\ref{solconst}).

Next we explain  how the fermionic degrees of freedom of the two theories are related. Let us define
\begin{equation}
\Omega = \frac{1}{\sqrt 2}+\frac{1}{\sqrt 2}\Gamma_{345}\ ,
\end{equation}
then $\Omega^2 = \Gamma_{345}$ and 
we see that
\begin{equation}
\Gamma_{012}\Omega = \Omega^{-1}\Gamma_{012}\ .
\end{equation}
A consequence of this is that if we define
\begin{equation}\label{epsilonrel}
\epsilon' = \Omega\epsilon\qquad \Psi' = l^{3/2}\Omega \Psi\ ,
\end{equation}
then
\begin{equation}
\Gamma_{012}\epsilon' = \epsilon'\qquad \Gamma_{012}\Psi' = -\Psi'\ ,
\end{equation}
and hence $\epsilon'$ can be thought of as parametrizying the supersymmetries preserved by an M2-brane along $x^\alpha$.  

The supersymmetry transformations now read
\begin{align}
\delta\Psi' &=  \Gamma^\alpha\Gamma^ID_\alpha X^I\epsilon' - \frac{1}{3!} \Gamma^{IJK}[X^I,X^J,X^K]\epsilon'\nonumber\\
\delta X^I &= i\bar\epsilon{\,'}\Gamma^I\Psi'\nonumber\\
\delta A_\alpha (\cdot)&=i \bar\epsilon{\,'} \Gamma_\alpha \Gamma^I[X^I,\Psi',\ \cdot\ ]\ .
\end{align}
These are exactly the variations of the maximally supersymmetric M2-brane model \cite{Gustavsson:2007vu,Bagger:2007jr}. Moreover, we see that the constraint (\ref{solconst}) for the field strength $F_{\alpha\beta}$ 
\begin{equation}
  F_{\alpha\beta}(\cdot) =-\varepsilon_{\alpha\beta\gamma}[X^I,D^\gamma X^I, \ \cdot\ ] - \frac{i}{2}\varepsilon_{\alpha\beta\gamma}[\bar\Psi',\Gamma^\gamma\Psi', \ \cdot\ ]\ ,
\end{equation}
is precisely the equation of motion for the field strength of the maximally supersymmetric M2-brane model. Similarly, the remaining equations of motion reduce to the correct equations of motion:\begin{align}
0=& D^2X^I+\frac{1}{2}[[X^I,X^J,X^K],X^J,X^K]+\frac{i}{2}[\bar \Psi',\Gamma^{IJ}\Psi',X^J]\nonumber\\
0=& \Gamma^\alpha D_\alpha \Psi'+\frac{1}{2}\Gamma^{IJ}[\Psi,X^I,X^J]\ .
\end{align}
Therefore we showed that upon imposing the solution of the constraints  (\ref{solconst}) on the (2,0) algebra (\ref{algebrafixed}) we obtain the maximally supersymmetric   model  describing two M2-branes. 

Let us briefly mention what happens if we instead take
\begin{equation}
C_{\alpha\beta\gamma} 
 = l^3\varepsilon_{\alpha\beta\gamma} \ .
\end{equation}
This is essentially just a double Wick rotation so that the equations are obtained by a suitable Wick rotation. Thus we arrive at a euclidean field theory in three dimensions. An inspection of the equations shows that this has an  $SO(2,6)$ R-symmetry.\footnote{One might object that the fermion $\Psi'$ is no longer real but there is no particular reason to make the redefinition from $\Psi$.} We thus obtain a non-abelian three-dimensional euclidean theory which is suitable to describe an euclidean M2-brane in $(5+6)$-dimensional spacetime, as appears in the work of \cite{Hull:1998fh}.

 \section{Conclusions}

 In this paper we have generalized the $(2,0)$ system of \cite{Lambert:2010wm} to include a background abelian three-form. 
The result is a maximally supersymmetric system of equations of motion with constraints whose solutions correspond to descriptions of M5-branes and M2-branes. Thus we have obtained a system of equations that furnish a representation of the six-dimensional $(2,0)$ superalgebra that plays an analogous role for M-branes that of ten-dimensional super-Yang-Mills does for D$p$-branes.

 The Lie 3-algebra here is known to have only one realisation with a positive definite invariant inner-product \cite{Papadopoulos:2008sk,Gauntlett:2008uf}. The corresponding M2-brane theory describes two M2-branes in ${\mathbb R}^8$ and ${\mathbb R}^8/{\mathbb Z}_2$ depending on whether or not one takes the gauge group $SO(4)$ and $Spin(4)$ respectively \cite{Lambert:2008et,Distler:2008mk,Lambert:2010ji,Bashkirov:2011pt}. The corresponding M5-brane equations of motion are then those associated to two M5-branes and gauge group $SU(2)$ along with a free centre of mass tensor multiplet. Thus the physical applications are somewhat limited but appear to capture all the known dynamics of two M2-branes or two M5-branes with maximal supersymmetry. Presumably, as with the case of M2-branes, the  case of two M-branes admits more manifest symmetries. Therefore it is hoped that there is a  broader description of M-branes that is valid for any number of branes. The extension to 12 supersymmetries and the ABJM model is currently under investigation.

 It would be interesting to understand the physical interpretation of $C_{\mu\nu\lambda}$. It is hard not to associate it with the bulk three-form of eleven-dimensional supergravity. For example in \cite{Ho:2008ve}, which had similar expression to what we derived in section 5,  $C_{\mu\nu\lambda}$ was viewed as giving rise to a non-commutativity. It would also interesting to understand the role of T-duality here and more generally in relating M2-branes and M5-branes.

It would be interesting to see if the system here has a natural interpretation in terms of higher gauge theory as in \cite{Palmer:2012ya}. We also note that although the system here leads naturally to M2-branes and M5-branes in the absence of $C_{\mu\nu\lambda}$ one can also obtain D-branes by considering non-positive definite 3-algebras as in \cite{Kawamoto:2011ab,Honma:2011br}. Therefore it would be interesting to explore the resulting system with non-vanishing three-form and non-positive definite 3-algebras. Finally there are other choices of 3-form that might lead to interesting new physical systems associated to M-branes.

\section*{Acknowledgements} 

N.L. is supported in part by STFC grant  ST/L000326/1, D.S. is supported by STFC grant ST/J0028798/1.
 
\bibliographystyle{utphys}
\bibliography{M2M5}

\end{document}